\documentclass[twocolumn,secnumarabic,amssymb, nobibnotes, aps, pre]{revtex4-2}

\usepackage{dcolumn}
\usepackage{bm} 
\usepackage{graphicx} 
\usepackage{color}
\usepackage{subfigure}
\usepackage{hyperref}
\usepackage{latexsym}
\usepackage{amsthm}
\usepackage{amssymb}
\usepackage{verbatim}
\usepackage{wasysym}

\usepackage{mathrsfs}

\DeclareGraphicsExtensions{.jpg, .pdf, .mps, .png, .eps, .ps, .EPS, .gif}
\begin{document}

\title{Percolation in two-species antagonistic random sequential adsorption in two dimensions}
 \author{Paulo H. L. Martins}
 \email{pmartins@fisica.ufmt.br}
 \affiliation{Instituto de F\'{\i}sica, Universidade Federal de Mato Grosso, Av.~Fernando Corr\^ea da Costa,~2367, Cuiab\'a, Mato Grosso, 78060-900,  Brazil}
 
 \author{Ronald Dickman}
 \email{dickman@fisica.ufmg.br}
 \affiliation{Departamento de F\'{\i}sica and National Institute of Science and Technology for Complex Systems, ICEx, Universidade Federal de Minas Gerais, C.P.~702, Belo Horizonte, Minas Gerais, 30123-970, Brazil}
 
\author{Robert M. Ziff}
\email{rziff@umich.edu}
\affiliation{Center for the Study of Complex Systems and Department of Chemical Engineering, University of Michigan, Ann Arbor, Michigan 48109-2800, USA}

\date{\today}

\begin{abstract}
We consider two-species random sequential adsorption (RSA) in which species A and B adsorb randomly on a lattice with the restriction that opposite species cannot occupy nearest-neighbor sites.  When the probability $x_A$ of choosing an A particle for an adsorption trial reaches a critical value $0.626441(1)$, the A species percolates and/or the blocked sites X (those with at least one A and one B nearest neighbor) percolate.  Analysis of the size-distribution exponent $\tau$, the wrapping probabilities, and the excess cluster number shows that the percolation transition is consistent with that of ordinary percolation.  We obtain an exact result for the low $x_B = 1 - x_A$ jamming behavior: $\theta_A = 1 - x_B +b_2 x_B^2+\mathcal{O}(x_B^3)$, $\theta_B = x_B/(z+1)+\mathcal{O}(x_B^2)$ for a $z$-coordinated lattice, where $\theta_A$ and $\theta_B$ are respectively the saturation coverages of species A and B.  We also show how differences between wrapping probabilities of A and X clusters, as well as differences in the number of A and X clusters, can be used to find the transition point accurately.  For the one-dimensional case a three-site approximation appears to provide exact results for the coverages.
\end{abstract}

\maketitle

\section{Introduction}

The random sequential adsorption (RSA) model has been widely studied and has numerous applications to physical problems of adsorption and reaction.  RSA is of interest in the study of adsorption on binary alloys~\cite{LoscarBorziAlbano06}, polymer adsorption~\cite{AdamczykRomiszowskiSikorskia08}, and protein adsorption \cite{TalbotTarjusVanTasselViot00}.In this model, particles of a certain shape fall uniformly over a surface, and adsorb (irreversibly) if and only if there is no overlap with a previously adsorbed particle.  Each realization of RSA begins with an empty surface and eventually reaches a {\it jammed} configuration in which overlaps are inevitable, so that no further deposition is possible \cite{Feder80}.  RSA is of interest in both continuous space and on regular lattices; RSA of unit segments on the line, also known as the ``car-parking problem," was introduced by R\'enyi \cite{Renyi58}.

Percolation theory itself has a broad range of applications, not only in physics and mathematics, but in materials science, ecology, biology, social sciences, and others~\cite{Flory39,SykesEssam64,StaufferAharony94}. Recent applications include vaccine allocation for achieving herd immunity~\cite{PenneyYargicSmolinThommesAnandBauch21} and radiation oncology~\cite{DimouArgyrakisKopelman22}.

Percolation of the objects deposited in RSA has also received much attention \cite{RamirezCentresVogelValdes19,FurlanDosSantosZiffDickman20,KunduPratesAraujo22}.  We note that, different from usual Bernoulli percolation, which involves a set of {\it mutually independent} random variables, percolation in RSA is an intrinsically nonequilibrium, history-dependent process. 

Here we consider a variation of RSA with two single-site species, A and B, such that A particles can be nearest neighbors (and similarly for B particles), but A and B particles cannot occupy neighboring sites.  
(These rules correspond to RSA of particles in a Widom-Rowlinson lattice gas~\cite{DickmanStell95}.)
If a site has at least one A and at least one B nearest neighbor, it is permanently blocked (X) for adsorption. The system reaches a maximum (jammed) coverage that depends upon the relative rates of A and B particles striking the surface.

Numerous other `AB' models exist in the literature, where A's and B's adsorb and may react and desorb if they are nearest neighbors, for example.  Those models generally do not reach a frozen state except for the complete coverage of the lattice by either A or B \cite{ZiffFichthorn86,Evans93}.  In the model studied here, there is no reaction and the system always reaches a frozen jammed state. 

Let $r_A$ ($r_B$) be the rates (per site) at which A (B) particles arrive at the surface. We define $x_A = r_A/(r_A+r_B)$ and $x_B = r_B/(r_A+r_B)= 1 - x_A$ as the probabilities that the next arriving particle be A or B, respectively.  
Note $x_A$ and $x_B$ are the probabilities of A or B deposition {\it attempts}, not the probabilities of successful adsorption.  As noted above, sites for adsorption attempts are uniformly distributed over the lattice.  Since the interactions amongst particles are invariant under the exchange of A and B, it is clear that the coverages satisfy $\theta_A (t; r_A, r_B) = \theta_B (t; r_B, r_A)$, where $\theta_{A(B)}$ denotes the fraction of sites occupied by A(B) particles (mean over realizations of a given lattice size $L$), whilst the fraction of blocked sites satisfies $\theta_X (t; r_A, r_B) = \theta_X (t; r_B, r_A)$.  Taking the limit $t \to \infty$, we have, for the jammed coverages,
$\theta_A (x_A) = \theta_B (1-x_A)$, and $\theta_X (x_A) = \theta_X (1- x_A)$. Analogous relations hold amongst the percolation thresholds.

\begin{figure}[t]
\includegraphics[width=1.0\linewidth]{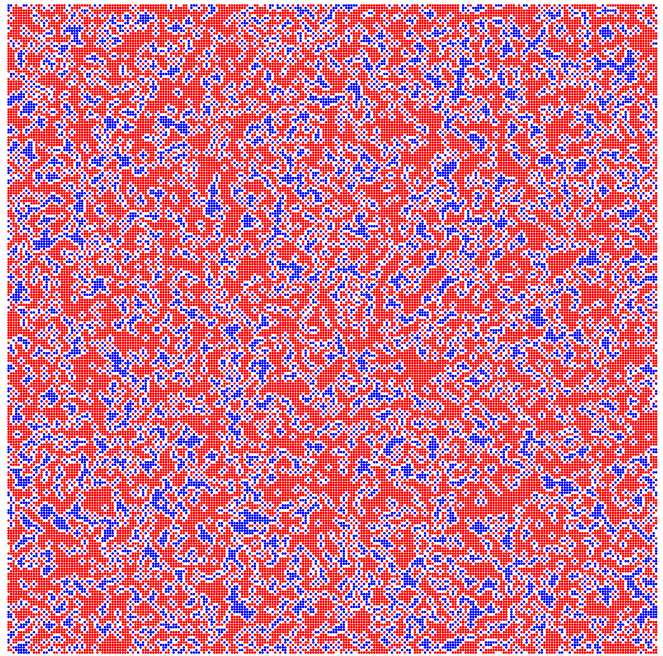}\\
\caption{A (red) and B (blue) sites at the critical point for A percolation: $x_A = 0.626441$ on a lattice of size 256$\times 256$. 
Blocked sites are represented by empty (white) spaces. }
\label{fig:ABrsa}
\end{figure}

\begin{figure}[!ht]
\includegraphics[width=1.0\linewidth]{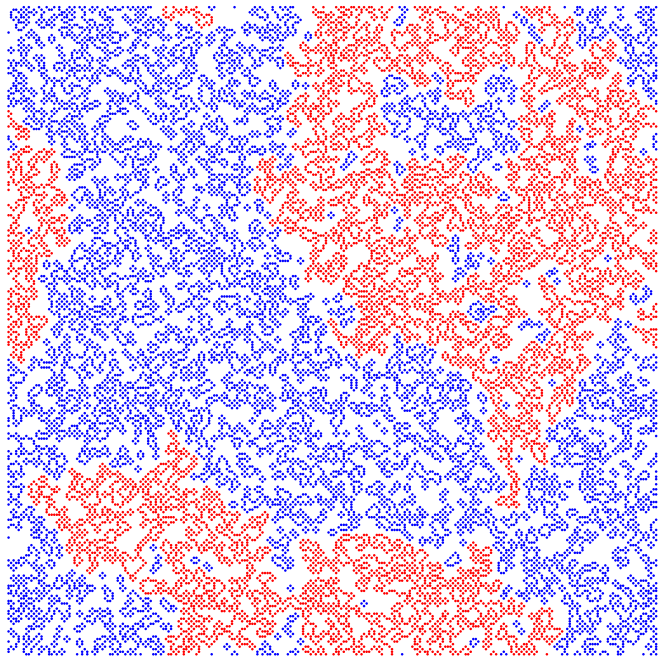}\\
\caption{Blocked sites (X) at the threshold, $x_A = 0.626441$, showing the largest X cluster in red, and all other X clusters in blue.  Sites occupied by A or B particles are shown in white.  }
\label{fig:ABrsablockedsites}
\end{figure}

Figures~\ref{fig:ABrsa} and \ref{fig:ABrsablockedsites} show typical jammed configurations on a 256$\times 256$ lattice at $x_A = 0.626441$, which corresponds to the critical threshold, as will be explained in Sec.~\ref{Sec:perc}. While Fig.~\ref{fig:ABrsa} shows A (red) and B (blue) particles, Fig.~\ref{fig:ABrsablockedsites} highlights the blocked (X) sites. The largest X cluster is highlighted in red and all other X clusters in blue, while sites with A or B adsorbed species are shown in white.  Note that X sites can form horizontal or diagonal neighbors, and occasionally make right angles and $2\times2$ squares, but cannot make larger solid blocks or assume a ``T" shape.  In general, the X sites form the boundaries between A and B clusters, and are therefore generally in the form of hulls. 

The remainder of this paper is organized as follows. In section II, we define the algorithm used in this study.  In section III, we discuss the jamming coverages, including an exact analysis of the small-$x_B$ behavior. In section IV we analyze percolation, including methods to obtain precise estimates for the thresholds, and study wrapping probabilities, the cluster-size distribution, and excess cluster number.   Section V presents results for the saturation coverages in the one-dimensional case.  Section VI contains our conclusions.

\section{Algorithms}

In the simplest (na\"ive) algorithm, a site is chosen randomly, and an A-adsorption attempt is performed with probability $x_A$, and a B-adsorption attempt otherwise.  The attempt succeeds if the chosen site is empty and has no nearest-neighbors occupied by the opposite species.
This algorithm will be rather inefficient because sites already occupied by a particle will be tried again, and it will require many trials to find the last empty (``O") sites.    
To know that we have reached the jammed state, we mark a site by $X$ if it has at least one A and at least one B nearest neighbor when visited, and stop the simulation when the number of empty sites, $n_0 = n - n_A - n_B - n_X$, is zero, where the $n_i$ are the numbers of sites in state $i$. 

A more efficient algorithm and nearly as simple to program employs a list of all empty (O) sites; the site $j$ for the next adsorption attempt is chosen at random from the list, and check for the presence of A- and B-particles at the nearest-neighbor sites.  If we find that site $j$ is blocked, we mark the site as X and remove it from the list by swapping the final entry of the list to the position 
 held by the site $j$, thereby shortening the  by one. If, by contrast, site $j$ is not blocked, we attempt A or B adsorption with the appropriate probabilities.   
If the attempted adsorption is successful, site $j$ is removed from the list.  If adsorption of the selected species is not allowed, site $j$ remains on the list of O sites, and may become occupied at a later trial, assuming the neighborhood does not change to prevent it. 
Although this algorithm is not maximally efficient in the sense that each site is visited only once, the extra work revisiting sites is small and the bookkeeping is simple.  We find the number of trials per site is about 1.41, independent of system size. As before, each realization of the RSA process halts when the list of O sites is empty. 

\begin{figure}[htb]
\includegraphics[width=1.0\linewidth]{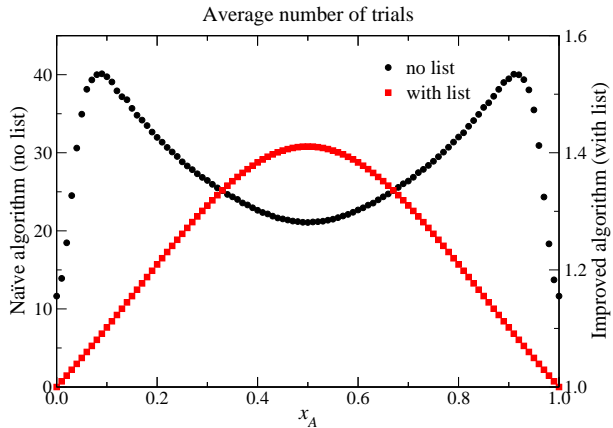}\\
\caption{Number of trials per site needed to saturate a $256 \times 256$ system. Note the different scales on the $y$-axis. Left labels correspond to the na\"ive algorithm (black dots), while right labels to the algorithm
employing a list of empty sites (red).}
\label{fig:ABrsaTrials}
\end{figure}

\begin{figure}[htb]
\includegraphics[width=1.0\linewidth]{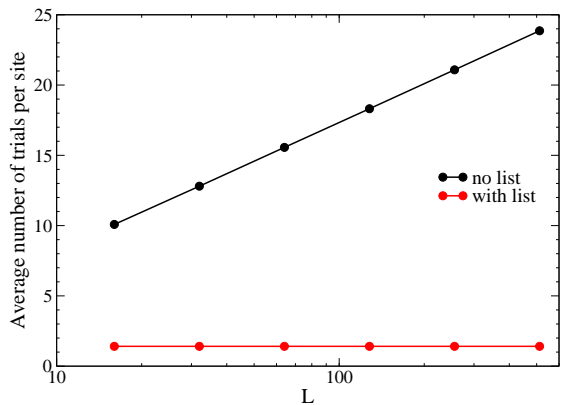}\\
\caption{Lattice size dependence of the average number of trials per site at $x_A=1/2$ (semi-log scale). Comparison between the na\"ive algorithm and that employing a list of empty sites.}
\label{fig:ABrsaTrialsL}
\end{figure}

The performance of the two algorithms is compared in Figs.~\ref{fig:ABrsaTrials} and \ref{fig:ABrsaTrialsL}.  For $x_A=1/2$ and $L=256$ for example, the na\"ive algorithm visits each site an average of $\approx 21$
times before the jammed state is reached. For other values of $x_A$, the number of trials is even higher, attaining a maximum of $\approx 40$ for $x_A \approx 0.09$. On the other hand, for the single-list algorithm, the number of trials increases monotonically from 1 at $x_A=0$ to $x_A=1/2$, for which the number of trials per site $\simeq 1.41$.

Figure~\ref{fig:ABrsaTrialsL} shows, in semi-log scale, the lattice size dependence of the average number of trials per site $T$ until saturation occurs, at $x_A=x_B=1/2$. For a system of size $L \times L$, we find that $T$ grows as $\ln L$, if the na\"ive algorithm is used, while $T$ remains independent of $L$ for the single-list algorithm.

Finally, a rejection-free algorithm can also be constructed in which one keeps two lists -- one of the sites in which an A particle can adsorb, the other of those at which a B particle can adsorb.  After an A is successfully adsorbed, the four neighbors are searched, and if any O are found, those sites are removed from the B adsorption list, and likewise for B adsorption.  However, we can also create an X site if after an adsorption, a neighboring O site is itself the neighbor of a site of the opposite species, and the X sites must be removed from both the A and B lists.  Since this method requires substantial bookkeeping following each adsorption, we opted for the single-list algorithm described above.

\section{Jamming coverages}

\begin{figure}[htb]
\includegraphics[width=1.0\linewidth]{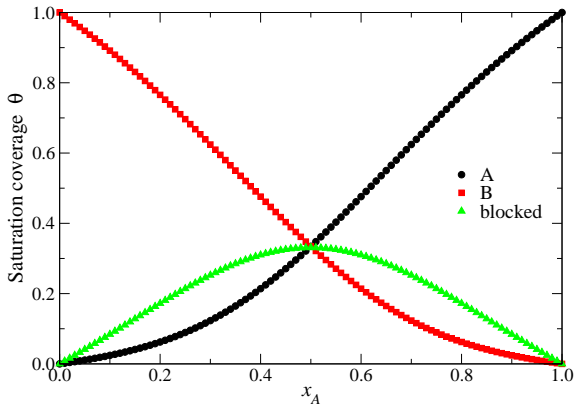}\\
\caption{Jammed coverages of species A and B, and of blocked sites X, on a 256$\times$256 lattice. 
}
\label{fig:ABrsaCoverages}
\end{figure}

The coverages of A, B, and X in the jammed state are plotted versus $x_A$ in Fig.~\ref{fig:ABrsaCoverages}.  For small $x_A$, the surface is mostly covered by B particles, as expected.  Significantly, however, as $x_A$ increases, the coverage of blocked sites grows faster than that of A particles.  In fact, as we show below, the number of X sites is four times the number of A sites in this regime.  It is also interesting to emphasize that for $x_A=x_B=1/2$, the coverage of X is slightly smaller than that of A or B, although this is not apparent at first sight in Fig.~\ref{fig:ABrsaCoverages}.  More precisely, we find $\theta_A=\theta_B=0.33451(1)$ and $\theta_X=0.33098(2)$.  
At the threshold, $x_A = 0.626441$, the coverages are $\theta_A=0.51516(1)$, $\theta_B=0.18695(1)$, and $\theta_X=0.29789(1)$, independent of $L$, to within uncertainty. That percolation occurs at a coverage far smaller than the threshold for Bernoulli site percolation on the square lattice, $p_c = 0.592746...$, \cite{Jacobsen15}, highlights the fact that occupation of nearby sites by the same species is highly correlated in the present model.
While we are unaware of exact expressions for the saturation coverages as functions of $x_A$, we describe below the behavior for small $x_B$. 

\subsection{Small-$x_B$ behavior}
\label{sec:smallp}
For small $x_B = 1 - x_A$, one expects the jamming coverages to follow power-series, i.e.,
\begin{eqnarray}
\theta_A &=& 1 - a_1 x_B - a_2 x_B^2 - \ldots \cr \cr
\theta_B &=& b_1 x_B + b_2  x_B^2 + \ldots \cr\cr
\theta_X &=& c_1 x_B + c_2 x_B^2 + \ldots 
\label{eq:thetasmall}
\end{eqnarray}
Since $\theta_A+\theta_B+\theta_X = 1$ at jamming, we have $a_1 = b_1 + c_1$ and $a_2 = b_2 + c_2$.

We carried out extensive simulations at $x_A = 0.995$, 0.996, 0.997, 0.998, and 0.999 on a square lattice of size 1024$\times$1024, with $3.5\times 10^6$ trials for each value of $x_A$. Fitting the results to a cubic polynomial in $x_B$ with intercept zero or one, we find $a_1 = 1.00012$, $a_2 = 0.708$, $b_1 = 0.20002$, $b_2 = 0.298$, $c_1 = 0.80009$, and $c_2 = 0.414$.  These results suggest that $a_1 = 1$, $b_1 = 1/5$, and $c_1 = 4/5$; in fact, we can prove that these values are indeed exact.

Consider a system with $x_B \ll 1$, and follow a single B among the A particles arriving at the surface, which is initially empty (all O sites).  In order for the B to adsorb, that site and the four neighboring sites must remain empty until the B reaches the surface---that is, no A can hit any of these five sites before the B arrives. Because $x_B$ is assumed to be very low, there will be no B sites nearby on the surface to block any of the A's from adsorbing, so any A arriving at one of these five sites before the central B will certainly adsorb.  

Think of the particles raining down on the surface at a fixed rate $r$ of one particle per time unit ($r=1$), and consider the possible adsorption of the B particle at a given instant $t$. The probability that a column of incoming particles is devoid of A particles is $e^{-\lambda}$, where $\lambda = x_A r t$ is the mean of the Poisson process, which corresponds to the expected number of A particles falling down until the instant $t$.  Now, we must have all five columns (below the B particle and its four neighbors) devoid of A particles, so the probability of adsorbing the B at time $t$ is 

\begin{equation}
\hbox{$\mathscr{P}$}(t) = e^{-5 t x_A}
\end{equation}
Clearly, the probability is greatest when $t = 0$, and decreases as $t$ increases.  Integrating this from $t = 0$ to $\infty$ to account for all possible deposition times of the first B, we obtain the total probability that a B will adsorb (given that B was chosen to land, which occurs with probability $x_B$):
\begin{equation}
    \int_0^\infty \hbox{$\mathscr{P}$}(t) dt = \int_0^\infty  e^{-5 t x_A} dt = \frac{1}{5 x_A}
\end{equation}
For small $x_B$, we have $x_A \approx 1$, showing that the probability that a single B will adsorb is $x_B/5$ and proving $b_1 = 1/5$.  Each adsorbed B site will cause the four neighboring sites (still empty) to become blocked, implying that $\theta_X = 4 \theta_B$ for small $\theta_B$, and therefore $c_1 = 4/5$.  Finally, it follows that $a_1 = b_1 + c_1 = 1$, as observed numerically.

The fact that $a_1=1$ has the interesting implication that for small $x_B$, $\theta_A \approx x_A$.  This result is not obvious because it says that the net number of A's on the surface is the same as $x_A$, even though the actual adsorption process is more complicated (some B's adsorb while most are blocked).    

\begin{figure}[htbp]
\includegraphics[width=1.0\linewidth]{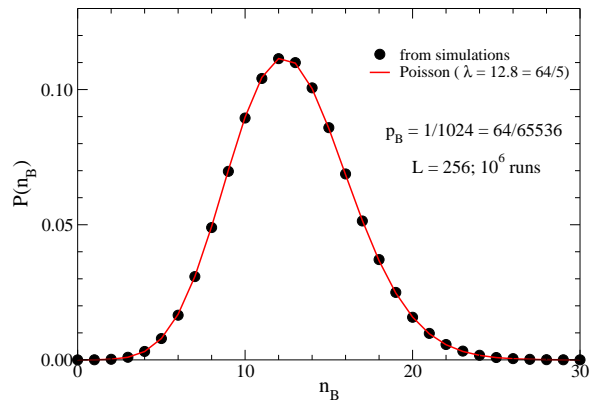}
\caption{Simulations for $L=256$ at $x_B=1-x_A=1/1024=64/L^2$; on average, for each $L^2$ adsorption attempts, a particle of species B is chosen $64$ times, leading to $64/5=12.8$ successful B adsorptions. The plot shows the probability of having $n_B$ B particles at saturation. Black dots show the results from simulations with $10^6$ runs each. The red line is a guide to the eye for the Poisson probability distribution with $\lambda = \langle n_B \rangle = 12.8$. }
\label{fig:ABrsaPoisson}
\end{figure}

To illustrate the adsorption behavior, we considered the case of an $L \times L$ lattice for $L=256$ with $x_B = 1/1024 = 64/L^2$.  For each of the $L^2$ attempts at adsorbing a particle, a B species is chosen $64$ times, on average. That corresponds to $64/5=12.8$ successful B landings. 
Figure~\ref{fig:ABrsaPoisson} compares the probability of having exactly $n_B$ adsorbed B particles at saturation (obtained from simulation) with the Poisson distribution with mean 12.8.  The agreement is excellent.  Thus, as one might expect, the overall B adsorption in the case of low $x_B$ is a Poisson process because for low $x_B$, the adsorption events can be considered independent.
For a lattice with coordination number $z$, by similar arguments it follows that $b_1 = 1/(z+1)$ and $c_1 = z/(z+1)$.  For example, for the triangular lattice, $b_1 = 1/7$.

The above results can also be derived by a simple argument, formulated here on the square lattice ($z=4$).  For small $x_B$, the chance that two B's strike the surface near each other is small, so we can focus on a single B striking the surface.  Consider the particles coming down on the surface above a central site and the four sites around it.  The probability that the first particle to strike the surface at the central site is a B is equal to $x_B$, and the probability that the four neighboring sites are all A's is $x_A^4$.  Now, the 5 particles can strike the surface in the order (B,A,A,A,A), (A,B,A,A,A), (A,A,B,A,A), (A,A,A,B,A), and (A,A,A,A,B), and only in the first case will the B adsorb, so the probability is 1/5.  (More precisely, we should multiply each by 4! for the permutation in the order of the A's landing, and then divide all by 5! permutations, so again we get 1/5 as the probability that a B lands before an A strikes one of the neighboring sites.)  As mentioned, for low $x_B$, we are assuming there are no nearby B's to forbid one of the A's to adsorb.

In summary, we find that when a single B particle is among the A particles striking a surface of size $L^2$, we get average numbers of each species adsorbed on an $L \times L$ system are given by
\begin{eqnarray}
n_A^{(1)} = L^2 - 1\cr\cr
n_B^{(1)} = 0.2\cr\cr
n_X^{(1)} = 0.8
\label{eq:oneB}
\end{eqnarray}
The last result follows from the fact that an adsorbed B particle will block all four nearest-neighbors from adsorption, and the results for $n_B^{(1)}$ and $n_X^{(1)}$ implies the result for $\theta_A$, since $n_A^{(1)}+n_B^{(1)}+n_X^{(1)}=L^2$.

\subsection{Second-order behavior}

We carried out simulations in which exactly two B particles attempt to adsorb at randomly chosen sites of an $L\times L$ system, with the rest of the particles As.  These two B particles are included in the list with the $(L^2-2)$ A particles, employing the list algorithm described in Sec.~II.  If a B does not successfully adsorb, then on the next random trial at that site, an A adsorption is attempted if the site is not blocked.

Fitting the results for $L = 32$, 64 and 128, we find for the expected numbers of each species adsorbed:
\begin{eqnarray}
n_A^{(2)} = L^2 - 2 -\frac{1.370}{L^2}\cr\cr
n_B^{(2)} = 0.4 +\frac{0.556}{L^2}\cr\cr
n_X^{(2)} = 1.6 +\frac{0.814}{L^2}
\label{eq:ni}
\end{eqnarray}
which adds up to $L^2$ as required.  The leading terms represent the results from Eq.\ (\ref{eq:oneB}) for independent adsorption, while the final terms reflect interactions (enhancement for B and X, depletion for A).

Now, we allow for different numbers of B's in the incoming stream, for low $x_B$.  The probability of choosing exactly $n$ B particles among the $L^2$ particles striking the surface is given by the Poisson distribution
\begin{equation}
    P(n) = \frac{\lambda^n}{n!} e^{-\lambda}
\end{equation}
with $\lambda = x_B L^2$ equal to the mean, which we assume to be small compared to 1.  Then we have for $\theta_i$, where $i =$ A, B, or X:
\begin{eqnarray}
    L^2 \theta_i &=& P(1) n_i^{(1)} + P(2) n_i^{(2)} +\ldots \cr \cr
    &=& x_B L^2 e^{-x_B L^2} n_i^{(1)} + \frac{(x_B L^2)^2}{2} e^{-x_B L^2} n_i^{(2)} +\ldots\cr \cr
    &=& x_B L^2 (1-x_B L^2) n_i^{(1)} +  \frac{(x_B L^2)^2}{2} n_i^{(2)}+\ldots\cr \cr
     &=& x_B L^2 n_i^{(1)} +  \frac{(x_B L^2)^2}{2} [n_i^{(2)}-2n_i^{(1)}] +\ldots
     \label{eq:thetai}
 \end{eqnarray}
up to order $x_B^2$.  Using our expressions for $n_B^{(1)}$ and $n_B^{(2)}$, we find for $\theta_B$
\begin{eqnarray}
    L^2 \theta_B &=& x_B L^2 (0.2) +  \frac{(x_B L^2)^2}{2} \frac{0.556}{L^2}\cr \cr
    &=& 0.2 x_B L^2 + 0.278 x_B^2 L^2
 \end{eqnarray}
 or 
 \begin{equation}
     \theta_B = 0.2 x_B + 0.278 x_B^2 +\ldots
     \label{eq:thetaB}
 \end{equation}
 which agrees fairly well with the simulation results of Eq.\ (\ref{eq:thetasmall}).

We can actually derive the expressions for $n_B$ in Eqs.\ (\ref{eq:ni}) directly and by considering the various cases where two B particles arrive to the surface close to each other.  By running our simulation with the relevant initial states, we can deduce the numerical values of the net adsorbed particles apparently exactly.

To begin, consider the vase in which the two B particles are set to arrive on nearest-neighbor sites; all remaining particles striking the surface are A's.  As above, if a B does not adsorb, and the site is not blocked, the next particle to strike that lattice site will be an A.  We find that the average number $n_i$ of particles of species $i$ to adsorb is equal to
\begin{eqnarray}
    n_A &=& L^2 - 2.3375 \cr
    n_B &=& 0.4925 \cr
    n_X &=& 1.8450
    \label{eq:BB}
\end{eqnarray}
In this case, the probability $p_B(n)$ that exactly $n$ B particles adsorb is given by $p_B(0)=0.5700$, $p_B(1)=0.3675$, and $p_B(2)=0.0625$, and the average number of B clusters $\langle N_B \rangle$ is equal to 0.4300.

Second, consider the case in which the two B particles are attempt to adsorb at a pair of second-neighbor
sites, e.g., $(0,0)$ and $(1,1)$.  Here we find:
\begin{eqnarray}
    n_A &=& L^2 - 2 \cr
    n_B &=& 0.43333 \cr
    n_X &=& 1.56666
    \label{eq:BBdiag}
\end{eqnarray}
Also $p_B(0)=0.6500$, $p_B(1)=0.26666$, and $p_B(2)=0.08333$, and $\langle N_B \rangle = 0.43333$, the same as $n_B$ because each adsorbed B is a distinct cluster.

Third, consider the case of the two B particles attempting to adsorb at third neighbors, e.g.,
$(0,0)$ and $(0,2)$; we find:
\begin{eqnarray}
    n_A &=& L^2 - 2 \cr
    n_B &=& 0.41111 \cr
    n_X &=& 1.58888
    \label{eq:BAB}
\end{eqnarray}
Here $p_B(0)=0.64444$, $p_B(1)=0.3$, and $p_B(2)=0.05555$, and $\langle N_B \rangle = 0.41111$, the same as $n_B$ because each adsorbed B is a distinct cluster.

When the two sites for attempted B adsorption are farther than third neighbors, we have simply
\begin{eqnarray}
    n_A &=& L^2 - 2 \cr
    n_B &=& 0.4 \cr
    n_X &=& 1.6
\end{eqnarray}

The above results Eqs.\ (\ref{eq:BB},\ref{eq:BBdiag},\ref{eq:BAB}) were found numerically to high precision and appear to be exact.  In fact, we have verified some of these results rigorously by arguments related to the probabilities of the relevant A's and two B's landing in the various combinatorial orderings, as an extension of the argument above proving that one B will adsorb with probability 1/5.

Now consider sending two B's randomly over the entire $L \times L$ system.  For the coverage of B, we get a term of $0.4 L^2$ assuming the B's are independent, and a correction arising from the 12 ways that the B's can be in the three patterns above.  (Fixing one of the Bs at the origin, there are four positions each where the second B can attempt to adsorb at a first, second, or third neighbor.)  For these cases, we subtract 0.4 because we want the difference from the independent result.  Thus we find,
\begin{equation}
    L^2 n_B^{(2)} = 0.4 L^2 + 4(0.0925+0.033333+0.011111)
\end{equation}
or,
\begin{equation}
    n_B^{(2)} = 0.4  + 0.54777/L^2
\end{equation}
in agreement with Eq.\ (\ref{eq:ni}).
Finally, inserting this result in Eq.\ (\ref{eq:thetai}), we find that $b_2 = 0.273888$, slightly lower than the fitted value $b_2 = 0.298$, which was rather approximate and depended significantly upon the order of the polynomial used to do the fitting.  We believe the value $b_2 = 0.273888 = 493/1800$ to be exact.

Similar results can be found for the cubic and higher-order terms in Eq.\ (\ref{eq:ni}), with somewhat more complicated considerations.  Results for $n_A$ and $n_X$ can also be found.

\section{Percolation}
\label{Sec:perc}

We considered percolation of A, B, and X clusters. A and B clusters are defined in the usual manner, i.e.,  percolation of nearest neighbors of the same species, while for X we allow both nearest and next-nearest neighbors, so that the coordination number for blocked-site clustersis $z=8$.  We consider the following events: wrapping in the horizontal direction $R^{(h)}$, in the vertical direction $R^{(v)}$, either $R^{(e)}$, both $R^{(b)}$, and in one direction but not in the other $R^{(1h)}$ and $R^{(1v)}$.  The theoretical values of these wrapping events on the torus at the critical point are known exactly~\cite{Pinson94,NewmanZiff00,NewmanZiff01}.

\begin{figure}[htbp]
\includegraphics[width=1.0\linewidth]{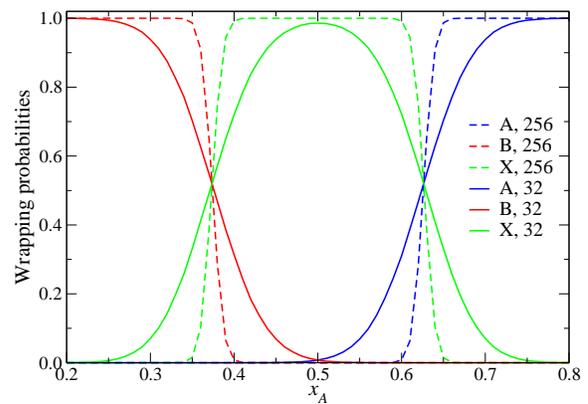}\\
\caption{Wrapping probabilities $R_i = (R_i^{(h)}+R_i^{(v)})/2$ of species A (in blue) and B (red), and of the blocked sites (green). Solid lines correspond to $L=32$ and dashed lines to $L=256$.}
\label{fig:ABrsaWrapping}
\end{figure}

Figure \ref{fig:ABrsaWrapping} shows the average wrapping probabilities $R_i = (R_i^{(h)}+R_i^{(v)})/2$ for two lattice sizes ($L=32$ and 256) at jamming. 
For $x_A=0$, where the surface is fully covered by B particles, one has $R_B=1$ and $R_A=R_X=0$. For small values of $x_A$, as discussed in Sec.~\ref{sec:smallp}, the number of blocked sites (X) is approximately 4 times the number of A particles. As $x_A$ increases towards 1/2, the coverage of blocked sites grows faster than the coverage of A particles (see Fig.~\ref{fig:ABrsaCoverages}), which results in a growing $R_X$. At the same time, $R_B$ decreases and becomes equal to $R_X$ at $x_A=0.373559$ or $x_B=0.626441$. This value corresponds to the percolation threshold of the B clusters, as will be explained further below. For $x_A = x_B = 1/2$, only a very small fraction of realizations yield percolating A or B clusters (i.e., clusters that wrap the lattice), while the wrapping probability for blocked sites is close to unity. Due to the symmetry mentioned in the Introduction, an equivalent behavior occurs for $x_A>1/2$, with the A percolation transition occurring at $x_A = 0.626441$. For larger lattices, the wrapping probability curves are steeper in the critical region, and in the limit $L\to\infty$ one expects a step function for $R_A$ and $R_B$ and a double step function for $R_X$.

We consider three different ways to determine the percolation threshold.  The first employs the wrapping probabilities, which are known to provide very accurate thresholds~\cite{NewmanZiff01}. In particular, if a given lattice (primary) is analyzed together with its corresponding matching lattice (the same lattice in which all polygons of more than three sides is made into a clique or complete graph), the threshold estimates converge extremely rapidly with increasing lattice size.  If only the primary lattice is used, one expects $|p_c(L) - p_c(\infty)| \sim L^{-2-1/\nu} = L^{-2.75}$~\cite{NewmanZiff01}, which is already a very rapid convergence. For some systems, however, considering both primary and matching lattices, one has $|p_c(L) - p_c(\infty)| \sim L^{-w}$, with $w=4$ or even higher~\cite{Jacobsen14,Jacobsen15,MertensZiff16}.  Here, $x_{A,c}$ takes the place of $p_c$. The matching lattice of the square lattice (with nearest neighbors) is the square lattice with next-nearest neighbors.  In the model considered here, we assume the blocked (X) sites can percolate through next-nearest neighbors ($z=8$), while clusters of A and B species are formed considering only nearest neighbors ($z=4$).  Thus, A and X form a matching pair near the transition point for A clusters, with the B clusters representing a kind of impurity, and B and X form a matching pair near the B cluster transition, with the A clusters forming an impurity.

\begin{figure}[htbp]
    \includegraphics[width=\linewidth]{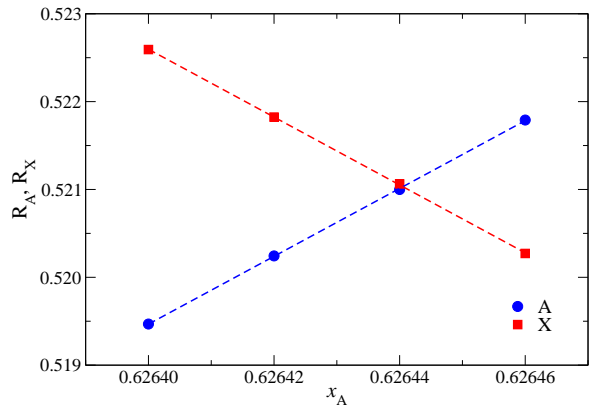}\\
    \caption{Percolation threshold obtained from the intercept between the horizontal wrapping probabilities 
    $R_A$ of A particles and $R_X$ of blocked sites for systems of size $L=256$, from a long run of $10^9$ samples.  The equations of the lines are  $R = -38.6231 x_A + 24.7161$ (upper) and $R =  38.6136 x_A - 23.6681$ (lower), and the intercept is listed in table \ref{tab:crossing}.}
    \label{fig:ABrsaCrossing}
\end{figure}

Figure~\ref{fig:ABrsaCrossing} shows our results for A and X wrapping. We performed extensive Monte Carlo runs in the neighborhood of the A critical point, requiring several months of CPU time.  As an example, for each value of $x_A$ shown in the plot, we averaged over $10^9$ realizations.  The mean wrapping probabilities $R_i(p)$ were obtained for A (in blue) and X clusters (in red). In this case, the intercept is located at $x_A=0.626441(1)$ and $R=0.521029(1)$ \cite{Pinson94,NewmanZiff01}.

The results for all $L$ are shown in Table \ref{tab:crossing}. The values of $x_A$ at the crossing points are independent of $L$ for $L \ge 64$ to the precision of our work, and imply a value $x_{A,c} = 0.626441(1)$.  The value of $x_A$ for $L = 32$ is only slightly higher (0.626445(1)), and suggests a very rapid convergence, such as $L^{-4}$.  Note that for smaller systems there will be occasional wrapping of B clusters which affects the results, so smaller systems should be used with caution to study the convergence of this estimate.

\begin{figure}[t]
\includegraphics[width=1.0\linewidth]{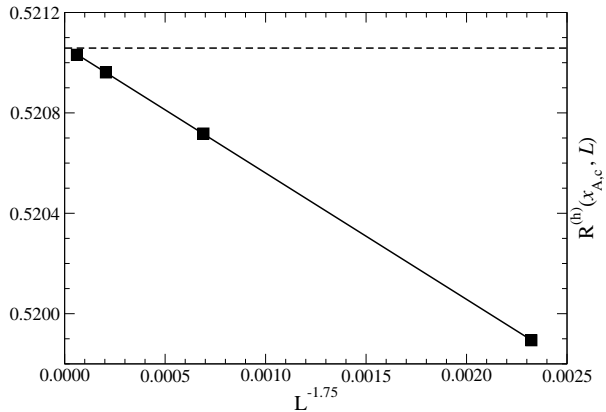}\\
\caption{Scaling of the wrapping probabilities at the critical point from Table \ref{tab:crossing}, vs.\ $L^{-1.75}$. The extrapolation gives $R_{\infty}(x_{A,c}) = 0.521064(10)$ in agreement with the exact value 0.521058. The exponent 1.75 was found empirically, and provides a much better fit than say exponent 2. }
\label{fig:ABrsaFSS}
\end{figure}

 In contrast to the rapid convergence of $x_{A,c}(L)$ to its asymptotic value as $L$ increases, the wrapping probabilities at the crossing point $R_A$ and $R_X$ approach their limiting values more slowly.  Figure~\ref{fig:ABrsaFSS} displays the scaling behavior of $R_A(x_{A,c},L)$, and is consistent with $|R_A(x_{A,c},L) - R_{\infty}(x_{A,c})| \sim L^{-1.75}$; this exponent, $1.75 = 1 + 1/\nu$ appears in some percolation convergences \cite{NewmanZiff01}. The intercept of the linear fit is consistent with the exact critical value $R_{\infty}(x_{A,c}) = 0.521058289$\cite{Pinson94,NewmanZiff01}, and a log-log plot of $(0.521058289-R_A(x_{A,c},L)$ vs.\ $L$ suggests an exponent of 1.78.
 This compares with an exponent of $\approx 2$ found in Ref.\ \cite{NewmanZiff01} for a similar quantity in site percolation (the wrapping probability at the critical point).

\begin{table}
\caption{Crossing points for $(R_i^{(h)}+R_i^{(v)})/2$ wrapping for A and X clusters, from plots like Fig.\ \ref{fig:ABrsaCrossing}.}
   \begin{tabular}{cccc}
   \hline\hline
        L & $N_\mathrm{runs}$ & $x_A$ & $R_A$ \\ \hline
        32 & $2\cdot 10^9$  & 0.6264449 & 0.5198968 \\ 
        64 & $2\cdot 10^9$  & 0.6264408 & 0.5207177 \\ 
        128 & $10^9$  & 0.6264407 & 0.5209613 \\ 
        256 & $10^9$ & 0.6264406 & 0.5210317 \\ 
       \hline
    \end{tabular}
    \label{tab:crossing}
\end{table}

\begin{figure}[ht]
\includegraphics[width=1.0\linewidth]{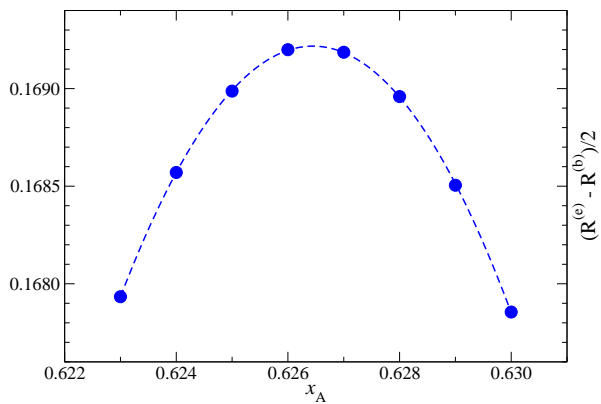}\\
\caption{Either minus both crossing divided by 2, equaling $(R^{(1h)}+R^{(1v)})/2$, as a function of $x_A$ for $L=64$.  The values at the maximum for different $L$ are listed in table \ref{tab:maxima}}.

\label{fig:ABrsaL128eitherboth}
\end{figure}

\begin{figure}[t]
\includegraphics[width=1.0\linewidth]{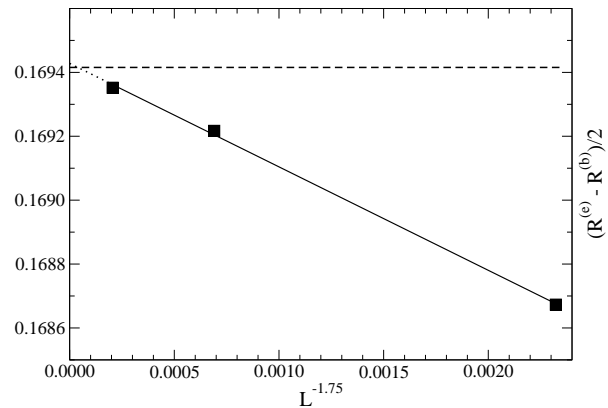}\\
\caption{Maxima of either minus both of A clusters (divided by 2) vs.\ $1/L^{1.75}$.  The intercept of the line at $L = \infty$ is $0.169428(14)$, consistent with the theoretical value for  is 0.16941544.}
\label{fig:ABrsaEitherBothMax}
\end{figure}

A second approach to finding the transition point is via the peak in $R_i^{(1h)}+R_i^{(1v)}=R_i^{(e)}-R_i^{(b)}$, the wrapping probability in one direction but not in the other one, equal to ``either" minus ``both" crossings.  It has the advantage that it peaks near the transition point, and it applies to individual species rather than having to consider the difference between two matching species such as A and X.  
However, the data near the peak show fairly large fluctuations, and in order to obtain a good parabolic fit we used a wider range in $x_A$ values, such as $x_A = 0.621$, $0.622,\ldots,$ 0.632 as shown in Fig.\ \ref{fig:ABrsaL128eitherboth} for $L = 128$. The values and locations of the peaks are shown in Table \ref{tab:maxima}; here a weak dependence of $x_A$ on $L$ is evident, but the precision is not sufficient to deduce the scaling. 
The value at the peak appears to approach the theoretical limiting value 0.16941544... as $L^{-1.75}$, as can be seen in Fig~\ref{fig:ABrsaEitherBothMax}.  Note an exponent of $-2$ is also possible here.

\begin{table}
\caption{Maxima of $(R^{(e)}-R^{(b)})/2=(R^{(1h)}+R^{(1v)})/2$, with a theoretical value of 0.169415435 \cite{Pinson94,NewmanZiff01}, from measurements at $x_A$ of values in the range $(0.621,0.630)$.}
   \begin{tabular}{cccc}
   \hline\hline
        L & $N_\mathrm{runs}$ & $x_A$ & $R_A$ \\ \hline
        32 & $2\cdot 10^9$  & 0.626484 & 0.168673\\ 
        64 & $2\cdot 10^9$  & 0.626447 & 0.169217 \\ 
        128 & $10^9$  & 0.626433 & 0.169352 \\
       \hline
    \end{tabular}
    \label{tab:maxima}
\end{table}

\begin{figure}[ht]
\includegraphics[width=1.0\linewidth]{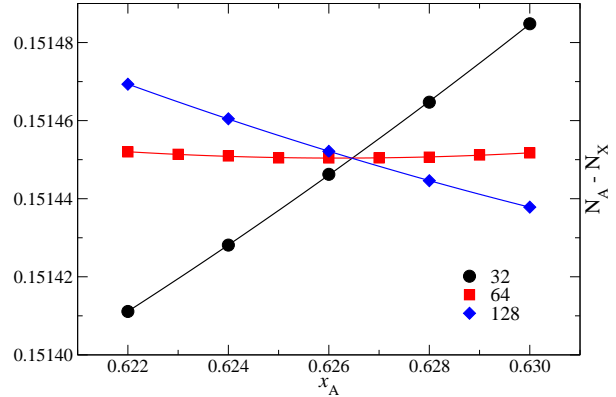}\\
\caption{Number of A clusters minus number of X clusters, plus $0.214 x_A$, added to make the slope of the middle curve close to zero for clarity, for $L=32$, 64, and 128.  The crossing points are listed in Table \ref{tab:clusters}, and all three  cross nearly at a common point $x_A \approx 0.62646$, $N_A-N_X\approx0.017388$.  Looking at cluster numbers is another way to find the percolation threshold for ordinary (uncorrelated) percolation \cite{MertensZiff16}, and it apparently is valid here as well.}
\label{fig:ABrsaNANXclusters}
\end{figure}

Finally, the third way that we can locate the threshold is via analysis of the density $N_A$ of clusters of A minus the density $N_X$ of clusters of X, using second-nearest-neighbor connectivity for X.  This quantity, on a pair of matching lattices, reaches an $L$-independent value, determined for regular lattices by the Euler characteristic $\chi(p)$, which here is the number of vertices, minus the number of edges, plus the number of elementary ($1 \times 1$) faces, per site \cite{SykesEssam64,MertensZiff16}.  For example, for site percolation on square lattices, $\chi(p_c) = p_c-2p_c^2 +p_c^4 = 0.0134956$~\cite{MertensZiff16} at $p_c=0.59274608$.  The procedure is depicted in Fig.~\ref{fig:ABrsaNANXclusters}, which shows  $N_A - N_X$ as a function of $x_A$ for three different lattice sizes.  The curves nearly cross at a common point, giving $x_{A,c} \approx 0.62645$, with $N_A - N_X \approx 0.01739$ at criticality.  As $N_A - N_X$ is a non-universal quantity, its value differs from the one given by the Euler characteristic for random percolation because of the correlations of the adsorbed particles.
The crossing points for some lattice sizes are listed in Table~\ref{tab:clusters}, and the values of $x_A$ are consistent with the threshold $x_{A,c}=0.626441$ determined by other methods. 

\begin{figure}[ht]
\includegraphics[width=1.0\linewidth]{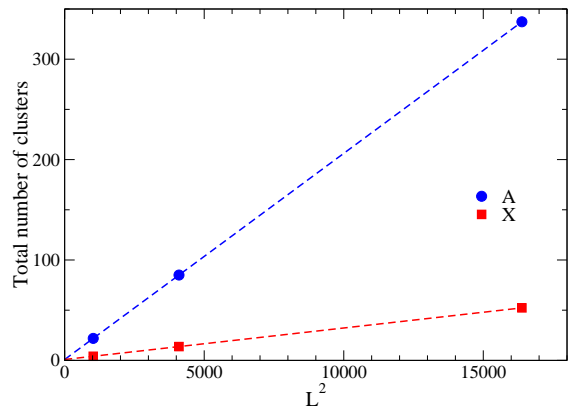}\\
\caption{ A plot of the number of A and X clusters vs.\ $L^2$ at criticality ($p_{A,c}$), both yielding an excess cluster number of 0.883 at the $y-$axis intercept (not really visible on this plot), consistent with the ordinary percolation universality class for a system with a square boundary.   }
\label{fig:ABrsaExcessNumber}
\end{figure}

\begin{figure}[t]
\includegraphics[width=1.0\linewidth]{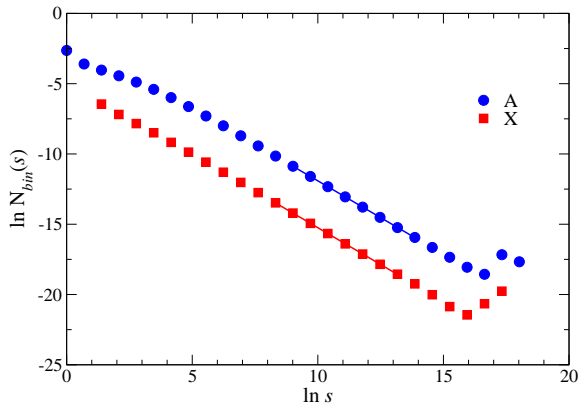}\\
\caption{Logarithm of the number of clusters with size in bins $(2^n,2^{n+1}-1)$ vs.\ $\ln s = \ln 2^n$, for A clusters (upper curve) and X clusters (lower curve) at $x_{A,c}=0.626441$ on a system of size 16384 $\times$ 16384 and $10^4$ runs.  For both species, the slope in the linear region (-1.046 for A and -1.051 for X) is close to the value for ordinary percolation  $1-\tau=-96/91=-1.055$.  Note the smaller finite-size corrections for small $s$ in the X-cluster case compared to the A-cluster case.}
\label{fig:ABrsaSizeDist}
\end{figure}

Having determined the percolation threshold, one can plot the total number of A and X clusters vs.\ $L^2$ at criticality, as shown in Fig.~\ref{fig:ABrsaExcessNumber}.  The slopes, which represent the density of clusters, yield $N_A=0.02053$ and $N_X=0.003138$ with $N_A - N_X \approx 0.01739$ independent of $L$.  At the threshold, the number of B clusters per site is $N_B = 0.057561$.  The $y$-axis intercept gives the excess number, which is universal for systems of a given shape, having the theoretical value 0.883576... for a square critical system \cite{ZiffFinchAdamchik97,KlebanZiff98}.  Our result of 0.883 supports the conclusion that the transition is in the ordinary percolation universality class.   Another confirmation of universality comes from the cluster size distribution at criticality, which behaves according $N{(s)} \sim s^{-\tau}$, where $s$ is the cluster size, $N{(s)}$ is the number of clusters (per site) of size $s$, and $\tau$ is the Fisher exponent ($\tau=187/91$ in two dimensions).  The  number of clusters within the range $(s,2s)$ is obtained by integration and behaves as $s^{-\tau+1}$. 
This behavior is clearly observed in Fig.~\ref{fig:ABrsaSizeDist}, which exhibits the logarithm of the number of clusters with size in bins $(2^n,2^{n+1}-1)$ vs.\ $\ln s = \ln 2^n$, for A and X clusters at $x_{A,c}=0.626441$ on a system of size 16384 $\times$ 16384 and $10^4$ runs.  For both species, the slope in the linear region is close to the ordinary percolation value $1-\tau=-96/91=-1.055$.  Note the smaller finite-size corrections for small $s$ in the X-cluster case.

At the A-cluster transition point, the B clusters act as impurities.  In fact, if we combine the B clusters with the X clusters, we find that the number of X clusters does not change---the B's are always entirely surrounded by X sites, and can simply be counted as part of the X clusters. Thus, the number of X clusters is the number of matching lattice clusters to the A clusters.  In Ref.\ \cite{MertensZiff16}, it is shown that for regular site percolation
\begin{equation}
    N(p) - N^*(p) = \chi(p) + f((p-p_c)L^{1/\nu})/L^2
\end{equation}
where $N(p)$ is the number of lattice clusters per site, $N^*(p)$ is the number of matching lattice clusters per site for lattice occupancy $p$, $\chi(p) = (\langle V \rangle - \langle E \rangle + \langle F_0)/L^2$ gives the average number of edges, vertices, and primal faces ($1\times1$ faces) in the graph connecting all nearest-neighbor occupied sites, and $f(z)$ goes from $-1$ to $+1$ as $z$ goes from $-\infty$ to $\infty$ with $f(0)=0$.  $f(z)$ is the wrapping probability on the lattice minus the wrapping probability on the dual lattice.  Expanding $f(z)$ by a Taylor series, we get
\begin{equation}
    N(p) - N^*(p) = \chi(p) + (p-p_c)L^{1/\nu-2}f'(0)
\end{equation}
Thus,  $N(p) - N^*(p)$ is a function that is weakly dependent upon $L$ and equal to $\chi(p_c)$ at $p = p_c$, for all $L$.  That is, all curves cross at the threshold point.  (We have ignored weak higher-order corrections that lead to small deviations in the crossing point.)

Analogously, we consider $N_A-N_X$, where as mentioned above $N_X$ is effectively $N_A^*$, shown in Fig.\ \ref{fig:ABrsaNANXclusters}.  (Here, we subtract a linear function to all the curves in order to make the crossing point more visible.) Indeed, we find a crossing point at the threshold $x_{A,c}= 0.62646$.  The value of $\chi$ at the threshold is 0.17388, different from the random value for site percolation, because here the number of vertices, edges, and primal faces is not simply related to $p$ as it is in random percolation.   Again, we find behavior consistent with percolation, but some different parameter values ($\chi(p_c)$) due to the correlations in the system. But in general the quantity $N_A-N_X$ behaves very much like the corresponding quantity (clusters minus matching-lattice clusters) in ordinary percolation \cite{MertensZiff16}---because the X and B together represent the matching system to the A clusters.

\begin{table}
\caption{Crossing points for the number of A clusters minus the number of X clusters, from the data shown in Fig.\ \ref{fig:ABrsaCrossing}.}
   \begin{tabular}{ccc}
   \hline\hline
        $L$  & $x_A$ & $N_A-N_X$ \\ \hline
        32 \& 64  & 0.626451& 0.017390\\ 
        64 \& 128 & 0.626469 & 0.017386\\ 
       \hline
    \end{tabular}
    \label{tab:clusters}
\end{table}

\section{One dimension}

We also carried out a study of the model in one dimension. Details will be given in another publication \cite{DickmanMartinsZiff22}; here we summarize some of the main results.

We consider a three-site approximation in which we focus on groups of neighboring sites $(\sigma_i,\sigma_j,\sigma_k)$ where $\sigma_i \in \{0,A,B\}$, where 0 is an empty site that may be blocked or not.  In order to write closed time-dependent rate equations for the three-site probabilities, it is necessary to write four-point probabilities  in terms of those for three points.  We use the factorization approximation
\begin{equation}
    p(\sigma_i,\sigma_j,\sigma_k,\sigma_\ell) \simeq \frac{ p(\sigma_i,\sigma_j,\sigma_k)p(\sigma_j,\sigma_k,\sigma_\ell)}{p(\sigma_j,\sigma_k)}
\end{equation}
where $p(.)$ is the time-dependent probability that the given events occur.  There are seventeen possible triplets: (000), seven combinations of 0 and/or A sites, an equal number of combinations of 0 and B, and the two mixed cases (A0B) and (B0A) in which the central sites are blocked.  The seventeen equations, including the above factorization approximation, are integrated using a fourth-order Runge-Kutta scheme, subject to the initial condition $p(000;t=0)=1.$  Taking $x_A = 1/2$, we found the final coverage of A and B particles of $\theta_\infty = 0.816060279$, which we recognize as simply $1-1/(2e)$.

As a second approach, we studied the jamming coverage for small periodic systems of length $L=4, 5, 6...$ to high precision at $x_A=1/2$ and found strong numerical evidence for the closed-form expression
\begin{equation}
    \theta_\infty(L) =  1 - \frac12  \sum _{k=0}^L \frac{(-1)^k}{k!} + \frac{L-1}{2L!}
\end{equation}
which also implies the limiting behavior $\theta_\infty = 1-1/(2e)$ for $L \to \infty$.  Note that the last two terms above, equal to  $(1-\theta_\infty(L))$, represent the number of ``derangements" of $L$ elements with an even number of cycles \cite{OEISA216778}, divided by $L!$.

\begin{figure}[t]
\centering
\includegraphics[width=1.0\linewidth]{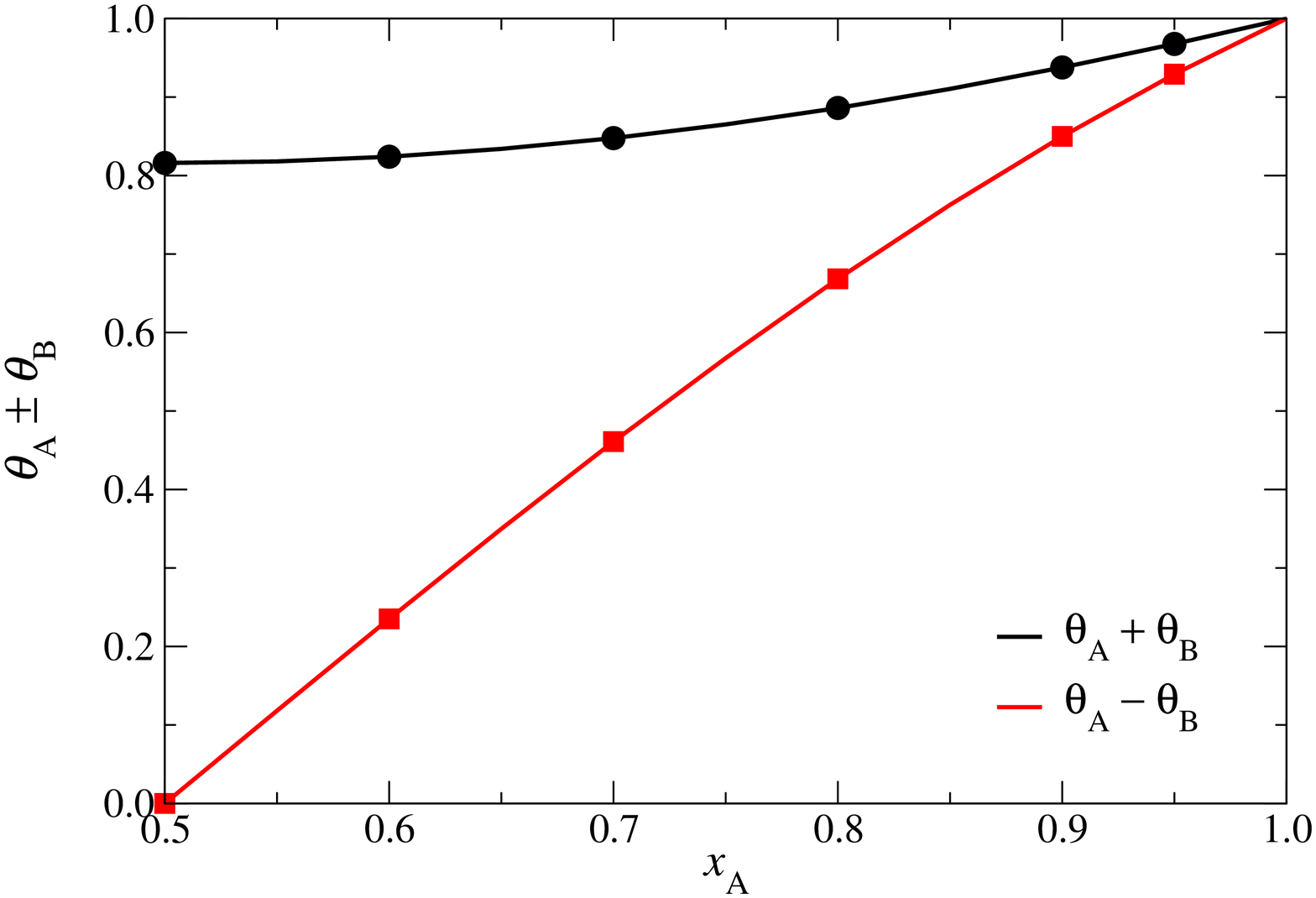}\\
\caption{Jamming coverage in one dimension, obtained from event-driven Monte Carlo simulations. }
\label{fig:ABrsa1dim}
\end{figure}

Finally, we carried out extensive event-driven Monte Carlo simulations on systems with $L$ up to 50000, using a two-list algorithm (those sites where an A can adsorb, and those sites where a B can adsorb), to find $\theta_\infty = 0.816060(1)$ at $x_A = 1/2$, in excellent agreement with the results of the above two methods. Figure~\ref{fig:ABrsa1dim} shows the total coverage for other values of $x_A$.

Thus, it appears that the three-site approximation is exact, in spite of the fact that there are undoubtedly correlations beyond that distance.  In one dimension there is always blocking, in that once a particle is adsorbed, the behavior on one side is independent of the behavior on the other site, and this can be used to study this system.   

We also studied the small-$x_B$ behavior and verified that $\theta_B=(1/3)x_B + \ldots$ consistent with the arguments in subsection \ref{sec:smallp} above for $z=2$.  Additional details of these results will be given in \cite{DickmanMartinsZiff22}.

\section{Conclusions}

In conclusion, we find that the AB RSA model shows intricate, interconnected adsorption and percolation behavior.  For RSA, we derive exact expressions for the coverages at low and high values of $x_A$, proving exactly the leading correction terms by a simple argument.   In terms of percolation, we show that the universality class is consistent with ordinary percolation, as one might expect because of the absence of long-range correlations.  There are two percolation transition points, one at $x_A = 0.626441(1)$ where A and X clusters percolate, and one at $x_A = 0.373559(1)$, where B and X clusters percolate.  We find the thresholds precisely from the crossing points of A (or B) wrapping, and X wrapping, and these predicted thresholds evidently show very small finite-size corrections.

We find that the total number of A clusters minus the number of X clusters, $N_A-N_X$, has a universal lattice-size-independent value 0.01739 at the transition point, consistent with the behavior found for ordinary percolation, although with a different crossing value because the occupation of the sites here is not random \cite{MertensZiff16}.  Here we have effectively a correlated model of percolation, and this study shows that the analysis based on cluster numbers can be applied to correlated percolation also.

The excess cluster number at criticality gives the identical value (0.883) as in ordinary percolation.  This is because the excess is related to large clusters in the system which are evidently unaffected by the correlations contained in the cluster growth process.
At the percolation transition, the cluster-size distribution also shows the behavior of standard percolation with exponent $\tau = 187/91$.

Possible directions for future study are series-expansion analyses for the time-dependent and jamming RSA coverages, behavior on different lattices and in different dimensions, and the transport properties of blocked-site clusters as models for porous media.
\vspace{1em}

{\bf Acknowledgements}
\vspace{1em}

P.M. thanks R.Z. for the hospitality at the University of Michigan.  R.D. is grateful to CNPq, Brazil for financial support under project number 303766/2016-6.

\end{document}